\documentclass[traditabstract]{aa}

\usepackage{graphicx}
\usepackage{txfonts}
\usepackage{natbib}
\usepackage{amssymb}
\usepackage{amsfonts}
\usepackage{amsbsy}
\usepackage{amsmath}
\usepackage{pdflscape}

\newcommand{\fb}{\boldsymbol{f}}

\newcommand{\gb}{\boldsymbol{g}}
\newcommand{\Cb}{\boldsymbol{C}}

\newcommand{\nb}{\boldsymbol{n}}

\newcommand{\ssb}{\boldsymbol{s}}

\newcommand{\xb}{\boldsymbol{x}}
\newcommand{\xbb}{\boldsymbol{\mathfrak{x}}}

\newcommand{\Erm}{\rm{E}}
\newcommand{\Tc}{\mathcal{T}}

\begin{document} 

\title{Correct estimate of the probability of false detection \\ of the matched filter  in weak-signal detection problems. III}
\subtitle{Peak distribution method versus the Gumbel distribution method}
 
   \author{R. Vio
         \inst{1}
          \and
          P. Andreani 
         \inst{2}
        \and
         A. Biggs
    \inst{2}
        \and
         N. Hayatsu
     \inst{2,3}
         }

   \institute{Chip Computers Consulting s.r.l., Viale Don L.~Sturzo 82,
              S.Liberale di Marcon, 30020 Venice, Italy\\
              \email{robertovio@tin.it}
          \and
                  ESO, Karl Schwarzschild strasse 2, 85748 Garching, Germany\\
                  \email{pandrean@eso.org}
             \email{pandrean@eso.org}    
              \and
                 Department of Physics Graduate School of Science, The University of Tokyo, 7-3-1 Hongo, Bunkyo, Tokyo 113-0033, Japan
           }      

   \date{Received....; accepted....}

\abstract{The matched filter (MF) represents one of the main tools to detect signals from known sources embedded in the noise. In the Gaussian isotropic case, the noise can be assumed to be the realization of a Gaussian 
random field (GRF). The most important property of the MF, the maximization of the probability
of detection subject to a constant probability of false detection or false alarm (PFA), makes it one of the most popular techniques. However, the MF technique relies upon the a priori 
knowledge of the number and the position of the searched signals in the GRF  (e.g. an emission line in a spectrum or a point-source on a map), which usually are not available.
A typical way out is to assume that, if present, the position of a signal coincides with one of the peaks in the matched filtered data. 
A detection is claimed when the probability that a given peak is due only  to the noise (i.e. the PFA) is smaller than a prefixed threshold. This last step represents a critical point in the detection procedure. 
Since a signal is searched for amongst the peaks, the probability density function (PDF) of the amplitudes of the latter has to be used for the computation of the PFA. Such a PDF, however, is different from the Gaussian. Moreover, the probability
that a detection is false depends on the number of peaks present in the filtered GRF. This is because the greater the number of peaks in a GRF, the higher the probability of peaks due to the noise
that exceed the detection threshold. If this fact is not taken into account, the PFA can be severely underestimated. In statistics this is a well-known problem named the multiple comparisons, multiple testing, or 
multiple hypotheses problem, whereas in other fields it is known as the look-elsewhere effect. Many solutions have been proposed to this problem. However, most of them are of a non-parametric type hence not able to exploit all the available information. 
Recently, this limitation has been overcome by means of two efficient parametric approaches. One is explicitly based on the PDF of the peak amplitudes of a smooth and isotropic 
GRF whereas the other makes use of the Gumbel distribution, which represents the asymptotic PDF of the corresponding extreme. On the basis of numerical experiments as well of an application to an interferometric map obtained with the
Atacama Large Millimeter/submillimeter Array (ALMA), we show that, although the two methods produce almost identical results, the first is more flexible and at the same time allows us to check the reliability of the detection procedure.}

   \titlerunning{Comparison of two methods for the computation of the probability of false alarm in signal detection problems}
   \authorrunning{Vio et al.}
   \maketitle
 
\section{Introduction} \label{sec:intro}

In essence, the matched filter (MF) is a linear low-pass filter that maximizes the signal to noise ratio of the detected signal. When the noise can be assumed as the realization of an isotropic Gaussian random field (GRF), this filter provides the greatest probability of detection for a fixed probability of false detection or false alarm (PFA) \citep{kay98}. This property makes the MF a very popular detection technique. However, MF requires the a
priori knowledge of the position of the signal within the GRF (e.g. an emission line in a spectrum or a point source in an
astronomical map). In most practical applications this is not the case. For this reason, the MF is used assuming that, if present, the position of a
signal corresponds to a peak in the matched filtered data. A detection can be assigned to the peaks exceeding a prefixed threshold. In other words, the detection procedure is not based on the entire area of the GRF but only on the subset of 
the points (pixels in the case of discrete data) corresponding to the position of a peak. Recently \citet{vio16} and \citet{vio17} (hereafter, VA16 and VVA17) have shown that, when based on the standard but wrong assumption that the probability density function (PDF) of the amplitudes of the peaks of a GRF is a Gaussian, this approach may lead to a severe underestimation of the PFA.  Moreover, the PFA does not provide
the probability that a given detection is spurious but  the probability that a generic peak can exceed, by chance, a fixed threshold. The critical  point is that the threshold 
has to depend on the number of peaks. The greater the number of peaks the higher the threshold. In statistics this is a well-known problem called the multiple comparisons, multiple testing or multiple hypotheses problem,
whereas in other fields it is known as the look-elsewhere effect. The majority of the
proposed solutions are essentially of a non-parametric type, such as the methods that control the family-wise error rate \citep{leh05} and the procedures to control  the false discovery rate \citep{ben95}, with the latter which has been proposed for the
astronomical applications \citep{mil01, hop02}. The main limitation of these methods is that, unlike the parametric approaches, they are not able to exploit all the available information. Parametric methods have been produced, but in general they are not
easy to use. An example is the procedure proposed by \citet{vit11}, based on the computation of the Euler 
characteristic of a GRF (a quantity more difficult to compute than the peak amplitudes), which needs numerical simulations to fix the value of some fundamental parameters.
For this reason, VA16 and VVA17 have introduced an efficient parametric approach able to provide a reliable estimate, 
called specific probability of false alarm (SPFA),
of the probability of a false detection. Such a quantity is computed on the basis of the PDF of the peak amplitudes of a smooth and isotropic  GRF\footnote{Actually, there are situations where the isotropy condition can be relaxed (see the last paragraph of Sect.~4 in Vio17).}. In a recent work, \citet{pav18} reach the same conclusions as VA16 and VVA17 but adopting a different approach based on the Gumbel distribution, which represents the asymptotic PDF of the extreme 
(i.e. the greatest value) of an isotropic GRF.

In this paper we show that, although the two approaches produce almost the same results, the method based on the PDF of the peak amplitudes (PAM) is more flexible than the method  based on the Gumbel distribution (GDM)  and moreover it permits an easy check of the conditions of its applicability. Hence, the detection procedure is more effective when PAM is used.

In Sect.~\ref{sec:MF1} the main characteristics of MF are reviewed. The reason why it underestimates the PFA when used in the standard way is explained in Sect.~\ref{sec:PAM}. In the same section,
the method suggested by VA16 and VVA17 
to correctly compute this quantity is illustrated  and the quantity SPFA introduced. The GDM approach is presented in Sect.~\ref{sec:GDM}.
Finally,  in Sects.~\ref{sec:comparison}-\ref{sec:ALMA} the PAM and GDM performances are tested on a set of simulated GRFs as well on an interferometric map obtained with the Atacama Large Millimeter/submillimeter Array (ALMA) and the final remarks are deferred to Sect.~\ref{sec:conclusions}.

\section{Matched filter} \label{sec:MF1}

Given a discrete observed signal $\xb$, the model assumed in the MF approach is
$\xb = \ssb + \nb$, where $\ssb$ is the deterministic signal to detect and $\nb$ a zero-mean GRF  with known covariance matrix,
\begin{equation} \label{eq:C}
\Cb = {\rm E}[\nb \nb^T]. 
\end{equation}
Here, symbols  ${\rm E}[.]$ and  $^T$ denote the expectation operator and the vector or matrix transpose, respectively.

Under these conditions,  according to the Neyman-Pearson theorem \citep{ney33, kay98}, a detection is claimed when
\begin{equation} \label{eq:test1}
\Tc(\xb) = \xb^T \fb > \gamma,
\end{equation}
with $\gamma$ a real constant and
\begin{equation} \label{eq:mf}
\fb_s = \Cb^{-1} \ssb.
\end{equation}
Here $\fb_s$ represents the matched filter. 
The main characteristic of the MF is that it maximizes the probability of detection under the constraint of a fixed PFA .

Matched filter  works properly under two assumptions. The first assumption is that the signal $\ssb$ is known. Actually, in practical applications only the template $\gb$ of the signal $\ssb = a \gb$ is available
but not its amplitude $a$. However, this does not represent a true problem since the  MF in the form
\begin{equation} \label{eq:mf2}
\fb_g = \Cb^{-1} \gb
\end{equation}
does not affect the PFA but only the probability of detection (VA16). The second assumption requires that $\xb$ and $\ssb$ have the same size. This implicitly means that the position of $\ssb$ within 
$\xb$  is known. Problems come up when this last assumption is relaxed and the sizes of $\ssb$ are smaller than the sizes of $\xb$ (e.g. an emission line in a spectrum). 
If the amplitude $a$ is also unknown, the standard approach consists in cross-correlating $\xb$ with the MF given by~Eq. \eqref{eq:mf2} obtaining signal $\xbb$ typically standardized to zero-mean and unit variance. 
A detection is claimed when a peak in $\xbb$ exceeds a threshold set to $u$.
If the number of signals $\ssb$ present in $\xb$ is also unknown, this procedure has to be applied to all the most significant peaks. It is a widespread practice that the corresponding PFA is given by
\begin{equation} \label{eq:fd1}
\alpha = \Phi_c(u),
\end{equation}
where  $\Phi_c(u) = 1 - \Phi(u),$ with $\Phi(u)$ the standard Gaussian cumulative distribution function (CDF).

\section{Detection procedure based on the PDF of the amplitude of the peaks} \label{sec:PAM}

In VA16 and VVA17  it has been shown that the computation of the PFA by means of Eq.~\eqref{eq:fd1} can lead to a severe underestimation of the latter quantity. This is because the PDF of the peaks of a GRF is not a Gaussian as
implicitly assumed in Eq.~\eqref{eq:fd1}. For this reason, the correct PFA has to be estimated by means of
\begin{equation} \label{eq:corra}
 \alpha = \Psi_c(u),
\end{equation}
where 
\begin{equation} 
\Psi_c(u)= 1 - \Psi(u),
\end{equation} 
with
\begin{equation} \label{eq:Psi}
\Psi(u)=\int_{-\infty}^{u} \psi(z) dz,
\end{equation} 
and
\begin{equation} \label{eq:pdf_z1}
\psi(z) = \frac{\sqrt{3 - \kappa^2}}{\sqrt{6 \pi}} {\rm e}^{-\frac{3 z^2}{2(3 - \kappa^2)}} + \frac{2 \kappa z \sqrt{\pi}}{\sqrt{6}} \phi(z) \Phi\left(\frac{\kappa z}{\sqrt{3 - \kappa^2}} \right)
\end{equation}
 for the 1D case, and
\begin{multline} \label{eq:pdf_z2}
\psi(z) = \sqrt{3} \kappa^2 (z^2-1) \phi(z) \Phi \left( \frac{\kappa z}{\sqrt{2 - \kappa^2}} \right) + \frac{\kappa z \sqrt{3 ( 2 - \kappa^2)}}{2 \pi} {\rm e}^{-\frac{z^2}{2 - \kappa^2}}\\
+\frac{\sqrt{6}}{\sqrt{\pi (3 - \kappa^2)}} {\rm{e}^{-\frac{3 z^2}{2 (3-\kappa^2)}}} \Phi\left( \frac{\kappa z}{\sqrt{(3 - \kappa^2) (2 - \kappa^2)}} \right)
\end{multline}
for the 2D case.
These expressions represent  the PDF of the local maxima of a zero-mean, unit-variance homogeneous GRF \citep{che15a, che15b} \footnote{For the three-dimensional case see  \citet{che15b}.}.

Here, $\kappa$ is a parameter given by
\begin{equation} \label{eq:kd}
\kappa = - \frac{\varrho'(0)}{\sqrt{\varrho''(0)}},
\end{equation}
where $\varrho'(0)$ and $\varrho''(0)$ are, respectively, the first and second derivative with respect to $r^2$ of the autocorrelation function $\varrho(r)$ of the GRF at $r=0$, with $r$ the inter-point distance of the random field.
As shown in VVA17, this parameter can be estimated by means of a maximum likelihood approach,
\begin{equation} \label{eq:ml}
\hat{\kappa} = \underset{\kappa }{\arg\max} \sum_{i=1}^{N_p} \log{\left(\psi(z_i; \kappa)\right)},
\end{equation}
where $\{ z_i \}$, $i=1,2,\ldots, N_p$, are the local maxima in $\xbb$  \footnote{We recall that the function ``$ \underset{x}{\arg\max}[ H(x)]$'' provides the value of $x$ for which the function $H(x)$ has the greatest value.}.
In this way, it is possible to avoid the use of Eq.~\eqref{eq:kd}, which requires knowledge of $\varrho(r)$.
The condition of validity of these expressions is that $\varrho(r)$ is differentiable at least six times with respect to $r$, but it is conjectured that four times should be sufficient (Cheng: private communication).
Especially in astronomy, this is a weak condition because due to the point spread function of the instruments, the MF often has a Gaussian-like shape. Hence, the resulting $\xbb$ are characterized by very smooth $\varrho(r)$.

It is necessary to stress that the PFA given by Eq.~\eqref{eq:corra} does not provide the probability $\alpha$ that a specific detection is spurious, but the probability that a generic peak due to the noise in $\xbb$ 
can exceed, by chance, the threshold $u$.
If the number of such peaks is $N_p$, then a number  $\alpha \times N_p$  among them  is expected to exceed the prefixed detection threshold. 
As a consequence, in spite of a low PFA, the reliability of a detection could actually be small. The solution proposed in VVA17 consists of a 
preselection based on the PFA and then in the computation of the specific probability of false alarm (SPFA) for each detection.
This quantity can be computed by means of the order statistics, in particular by exploiting  the statistical characteristics of the greatest value of a finite sample of identical and independently distributed (iid) random variable from a given 
PDF  \citep{hog13}. Under the iid condition, the PDF $\upsilon(z_{\max})$ of the largest value among a set of $N_p$ peaks $\{ z_i \}$ is given by
\begin{equation} \label{eq:gz}
\upsilon(z_{\max}) = N_p \left[ \Psi(z_{\max}) \right]^{N_p-1} \psi(z_{\max}).
\end{equation}
Hence, the SPFA can be evaluated by means of
\begin{equation} \label{eq:intz}
\alpha = \int_{z_{\max}}^{\infty} \upsilon(z') dz'.
\end{equation}
The numerical evaluation of this integral does not present particular difficulties since
\begin{align}
\alpha & = N_p \int_{z_{\max}}^{\infty} \left[\Psi(z) \right]^{N_p-1} d\Psi(z); \\
          & = \left[ \Psi(z) \right]^{N_p} \Big|_{z_{\rm max}}^{\infty}; \\
          & = 1-\Upsilon(z_{\rm max}),
\end{align}
with
\begin{equation} \label{eq:Ups}
\Upsilon(z_{\rm max})=\left[ \Psi(z_{\rm max}) \right]^{N_p}.
\end{equation}
This procedure is implicitly based on the assumption that there is only a signal $\ssb$ in $\xb$. If the actual number is unknown, it has to be cyclically applied to all the remaining most prominent peaks in order of decreasing amplitude and stopped
when the estimated PFA is greater than a prefixed $\alpha^*$. In principle, $N_p$ should be lowered by one unit after any detection. This is because $N_p$ represents the number of peaks due to the noise.
However, since in the practical applications $N_p$ is on the order of thousands if not tens of thousands, this step is irrelevant.

\begin{landscape}
   \begin{figure}
        \resizebox{\hsize}{!}{\includegraphics{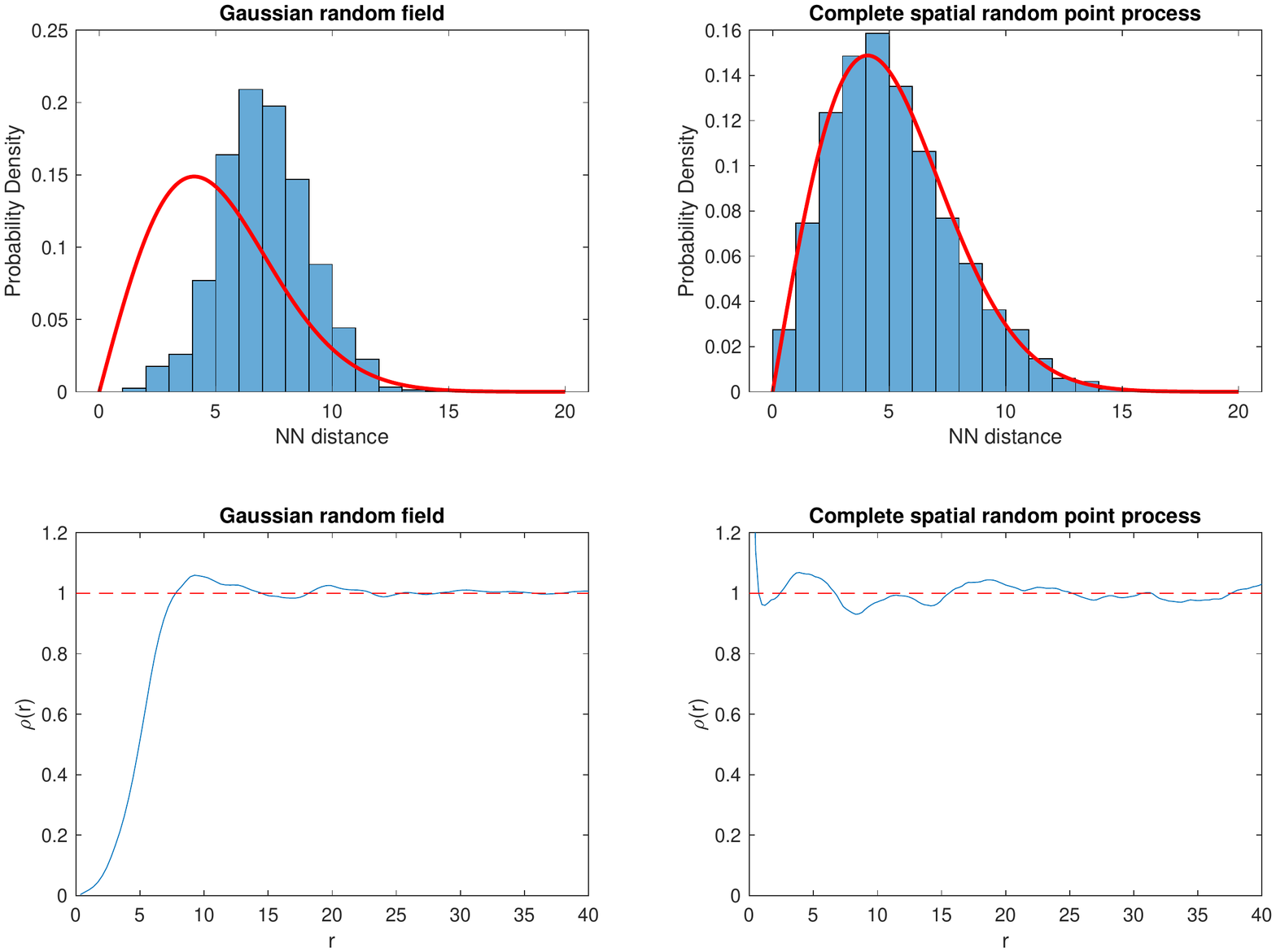}}
        \caption{Statistical characteristics of the spatial distribution of a set of points obtained by means of a GRF and a complete spatial random point process (CSRPP).
Top-left panel: Histogram of nearest neighbor distances of the peaks of a  simulated zero-mean unit-variance GRF with size $500 \times 500$ pixels and autocorrelation function given by a circular Gaussian 
with dispersion set to three pixels. Since in this GRF the peaks have coordinates given by integer numbers, in order to mimic a continuous spatial distribution, they have been added to a uniform random number
taking its value in the range $(-0.5, +0.5]$. Bottom-left panel: Sample pair correlation function $\rho(r)$ of the peaks in the same GRF. 
Top-right and bottom-right panels: Figures corresponding to the right panels for a CSRPP with the same sizes and containing a number of points equal to the number of peaks as in the GRF. 
The red lines provide the corresponding theoretical PDFs   due to a CSRPP.}
        \label{fig:fig_distance}
    \end{figure}
\end{landscape}

\begin{landscape}
   \begin{figure}
        \resizebox{\hsize}{!}{\includegraphics{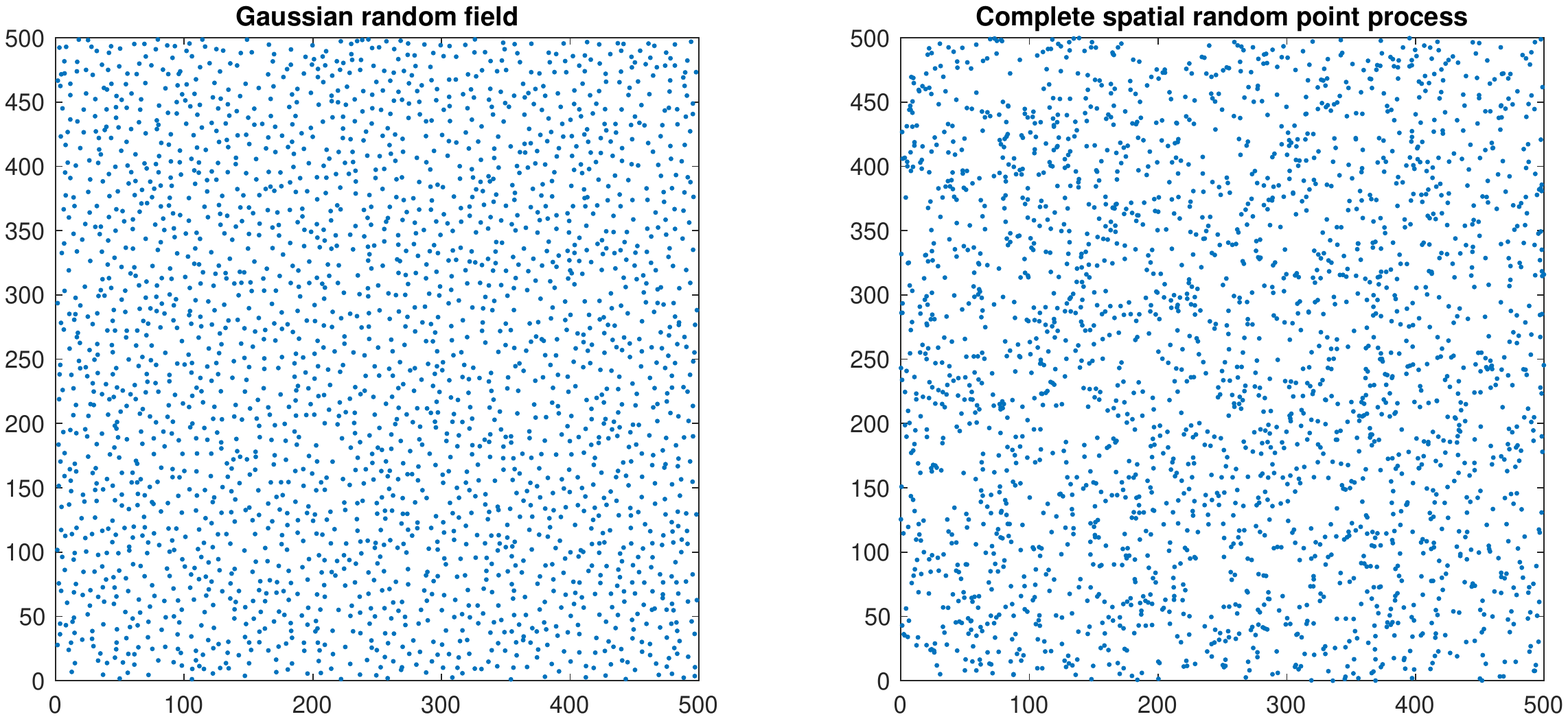}}
        \caption{Numerical simulation of the two processes used in Fig.~\ref{fig:fig_distance}.}
        \label{fig:fig_maps}
    \end{figure}
\end{landscape}

As a final note, it is necessary to stress that the peaks  of a isotropic GRF, with $\varrho(r)$ typical of many astronomical observations, usually have a spatial distribution different from the spatial pattern characteristic
of a complete spatial random point process (CSRPP).
This is visible in the top panels of  Fig.~\ref{fig:fig_distance} where the PDF of the nearest neighbor
distances of the peaks of a simulated GRF with $\varrho(r)$ given by a two-dimensional circular Gaussian (typical of many astronomical images) with dispersion $\sigma_G=3$ in pixel units is compared with the corresponding 
PDF of a simulated CSRPP.  From this figure it is clear that the PDF related to the peaks lacks
small values. Hence, the two processes are different on short spatial scales. This means that the iid condition for the peak amplitudes is not necessarily valid.
However,  on greater spatial scales, when $\varrho(r)$ is narrower than the
area spanned by the data (a basic situation for the application of the MF), this condition can be expected to hold with good accuracy. The rationale is that two generic points of a GRF with a distance $r$ such that $\varrho(r) \approx 0$ 
are essentially independent. The same holds for two generic peaks. This is confirmed by the bottom-left panel of Fig.~\ref{fig:fig_distance}  where the sample pair correlation function\footnote{The pair correlation function $\rho(r)$
of the spatial distribution of a set of points is given by
$\rho(r) = K'(r)/2 \pi r$ with $K'(r)$ the derivative of the Ripley's $K$-function with respect to $r$. For a CSRPP it is $\rho(r) = 1$ independently of $r$ \citep{bad16}.}  $\rho(r)$
of the peaks indicates that, for $r \ge 7,$ their spatial distribution is compatible with a  CSRPP.
For comparison, the bottom-right panel displays $\rho(r)$ for a true CSRPP.
The different behavior on small scales makes the peaks of the GRF show a more uniform spatial distribution than the spatial distribution of the points of the CSRPP (see Fig.~\ref{fig:fig_maps}).
In conclusion,  since  $\varrho(7) \approx 0$, it means that, for $r \ge 7,$ the peak amplitudes can be considered iid. As a result, most of the peak amplitudes of a GRF with $\varrho(r)$ typical of many astronomical applications (essentially Gaussian-like) can be expected to be approximately  iid. Hence,  Eq.~\eqref{eq:gz} is still applicable but possibly with an 
effective number smaller than $N_p$ \citep[see Sect. 6 in][]{may05}.
This last point is due to the dependence among a set of random variables, which lowers their number of degrees of freedom\footnote{The term degrees of freedom refers to the number of
items that can be freely varied in calculating a statistic without violating any constraints.}.

\section{Detection procedure based on the Gumbel distribution} \label{sec:GDM}

Independently of their PDF, the CDF of the greatest value $x_{\rm max}$ of a set of $N \longrightarrow \infty$ iid random variables $\xb$ is given by
\begin{equation} \label{eq:gumb}
G(x_{\rm max})=\exp{\left[ -(1+\gamma_g y)^{1/\gamma_g} \right]}
,\end{equation}
where 
\begin{equation}
y=\frac{x_{\rm max}-a}{b}
\end{equation}
with $a$ and $b$ the location and the scale parameter, respectively. Three cases are possible, according to  $\gamma_g < 0$, $\gamma_g=0$ and $\gamma_g > 0$ \citep{cas04}. Assuming that $x_{\rm max}$ corresponds to a peak, 
that is, $x_{\rm max}=z_{\rm max}$, and in the case of an isotropic GRF, it has been shown \citep{col11}
that the CDF of $z_{\rm max}$ is given by Eq.~\eqref{eq:gumb} with $\gamma_g =0$,
\begin{equation} \label{eq:gumb0}
G(z_{\rm max})=\exp{\left[ - {\rm e}^{-y} \right]},
\end{equation}
which represents the Gumbel distribution. With a two-dimensional, zero-mean, unit-variance GRF and in the regime  of slight clustering of the peaks, Eq.~\eqref{eq:gumb0} can be approximated by
\begin{equation}
G(z_{\rm max}) \approx \exp{\left[- \frac{\gamma^2 A z_{\rm max} }{(2 \pi)^{3/2} R_*^2} {\rm e}^{-z_{\rm max}^2/2} \right]},
\end{equation} 
where $A$ is the area of the GRF,
\begin{equation}
\gamma=\frac{\sigma_1^2}{\sigma_0 \sigma_2},
\end{equation}
and
\begin{equation}
R_*=\frac{\sigma_1}{\sigma_2} \sqrt{2}.
\end{equation}
The SPFA for a specific $z_{\rm max}$ is given by $\alpha=1-G(z_{\rm max})$.

If the GRF is obtained by filtering a noise process characterized by a scale-free power spectrum $P(k)$,
\begin{equation}
P(k) = B k^n,
\end{equation}
with a filter $w_l(k)$ ($l$ a scale parameter), it is
\begin{equation} \label{eq:sigma_i}
\sigma_j^2= \frac{1}{2 \pi} \int_0^{\infty} k^{2 j +1} P(k) w_l^2 (k) dk.
\end{equation}
Details of the three-dimensional case can be found in \citet{col11}.

\section{Comparison with simulated data} \label{sec:comparison}
 
Given the different derivation of PAM and GDM, it is necessary to compare their performances. With this aim, a numerical experiment was carried out that is based on $5 x 10^3$ 
numerical simulations of $500 \times 500$ pixels GRFs characterized by a  circular Gaussian autocorrelation function with dispersion $\sigma_G=3$.
This kind of GRF corresponds to what is obtained by filtering a two-dimensional white-noise with a circular Gaussian $w_l(x)$ with standard deviation $l=3/\sqrt{2}$. The resulting $G(z_{\rm max}) $ can be expressed in the analytical form
\begin{equation} \label{eq:GG}
G(z_{\rm max}) \approx \exp{\left[- \frac{r}{4 \sqrt{2 \pi}}  z_{\rm max} {\rm e}^{-z^2_{\rm max}/2}  \right]},
\end{equation}
where $N_*$ is the ratio of the total area of the GRF to the window area $\pi  l^2$ \citep{pav18}. 

\begin{figure*}
        \resizebox{\hsize}{!}{\includegraphics{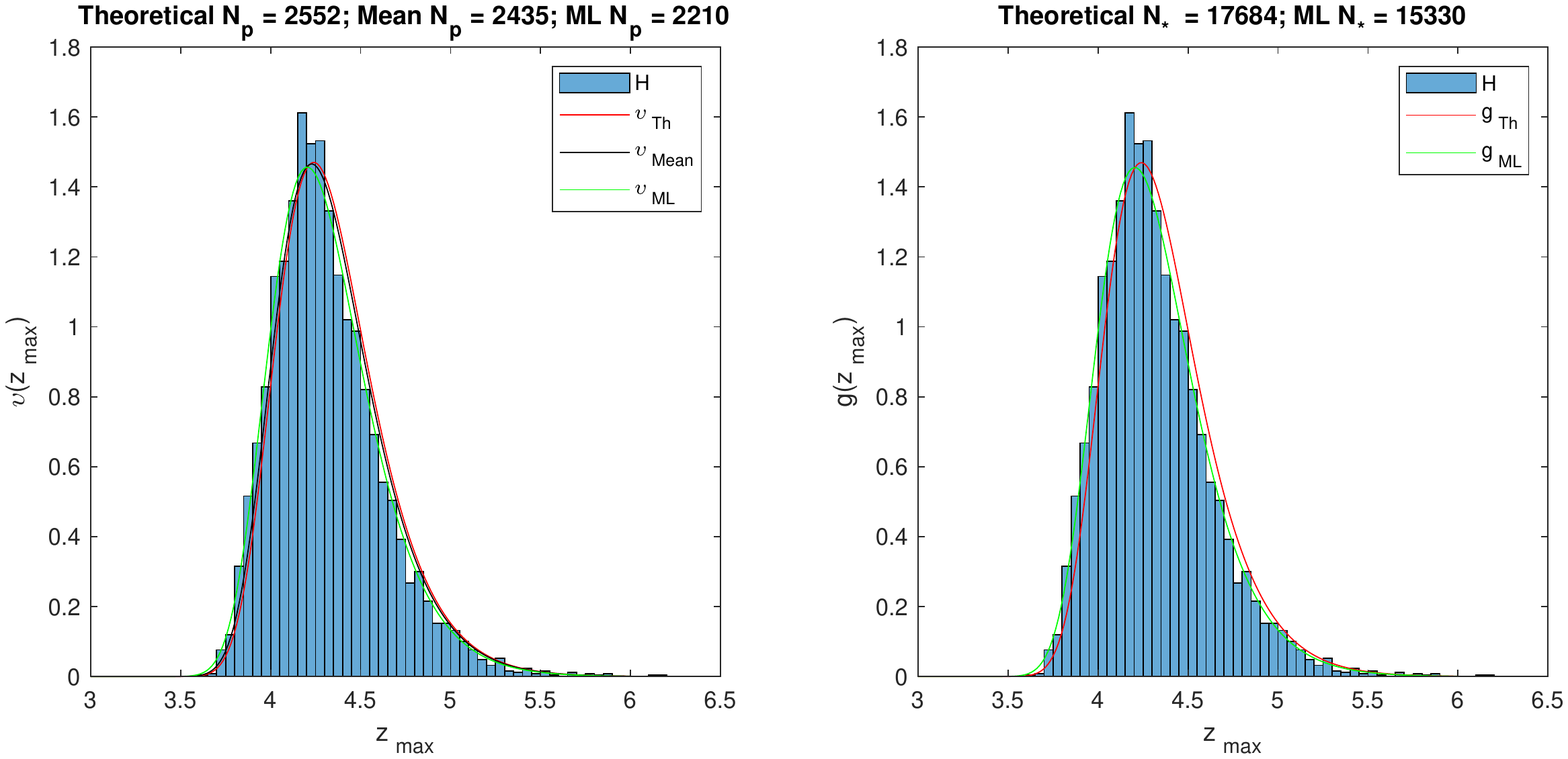}}
        \caption{Left panel: histogram $H(z_{\rm max})$ vs. the theoretical PDF $\upsilon_{\rm Th}(z_{\rm max})$, mean PDF $\upsilon_{\rm Mean}(z_{\rm max})$ and the maximum likelihood PDF $\upsilon_{\rm ML}(z_{\rm max})$
of the value $z_{\rm max}$ of the highest peak of a zero-mean unit-variance GRF with autocorrelation given by a circular Gaussian with dispersion set to three pixels. The numerical experiment is based on the simulation  
of $5 \times 10^3$ GRF's of size $500 \times 500$ pixels. $\upsilon_{\rm Th}(z_{\rm max})$ has been computed using Eqs.~\eqref{eq:Ups} and \eqref{eq:Psi} with $N_p$ obtained as explained in Vio17, whereas 
$\upsilon_{\rm mean}(z_{\rm max})$  has been computed with $N_p$ given by the mean number of peaks in the simulated GRF's. In the case of $\upsilon_{\rm ML}(z_{\rm max})$ the number $N_p$ comes out from a maximum likelihood method
(see text). Right panel: corresponding $g_{\rm Th}(z_{\rm max})$, computed by means of Eq.~\eqref{eq:GG} with $N_*$ given by  the ratio of the total area of the GRF to the window area $\pi  l^2$ with $l=3/\sqrt{2}$ (see text). Also the PDF $g_{\rm ML}(z_{\rm max})$ has been computed with $N_*$ resulting from a maximum likelihood method. To notice that $\upsilon_{\rm Th}(z_{\rm max})$ and $g_{\rm Th}(z_{\rm max})$ are almost identical. The same holds for 
$\upsilon_{\rm ML}(z_{\rm max})$ and $g_{\rm ML}(z_{\rm max})$.}
        \label{fig:fig_pdf_circ}
    \end{figure*}

\begin{landscape}
   \begin{figure}
        \resizebox{\hsize}{!}{\includegraphics{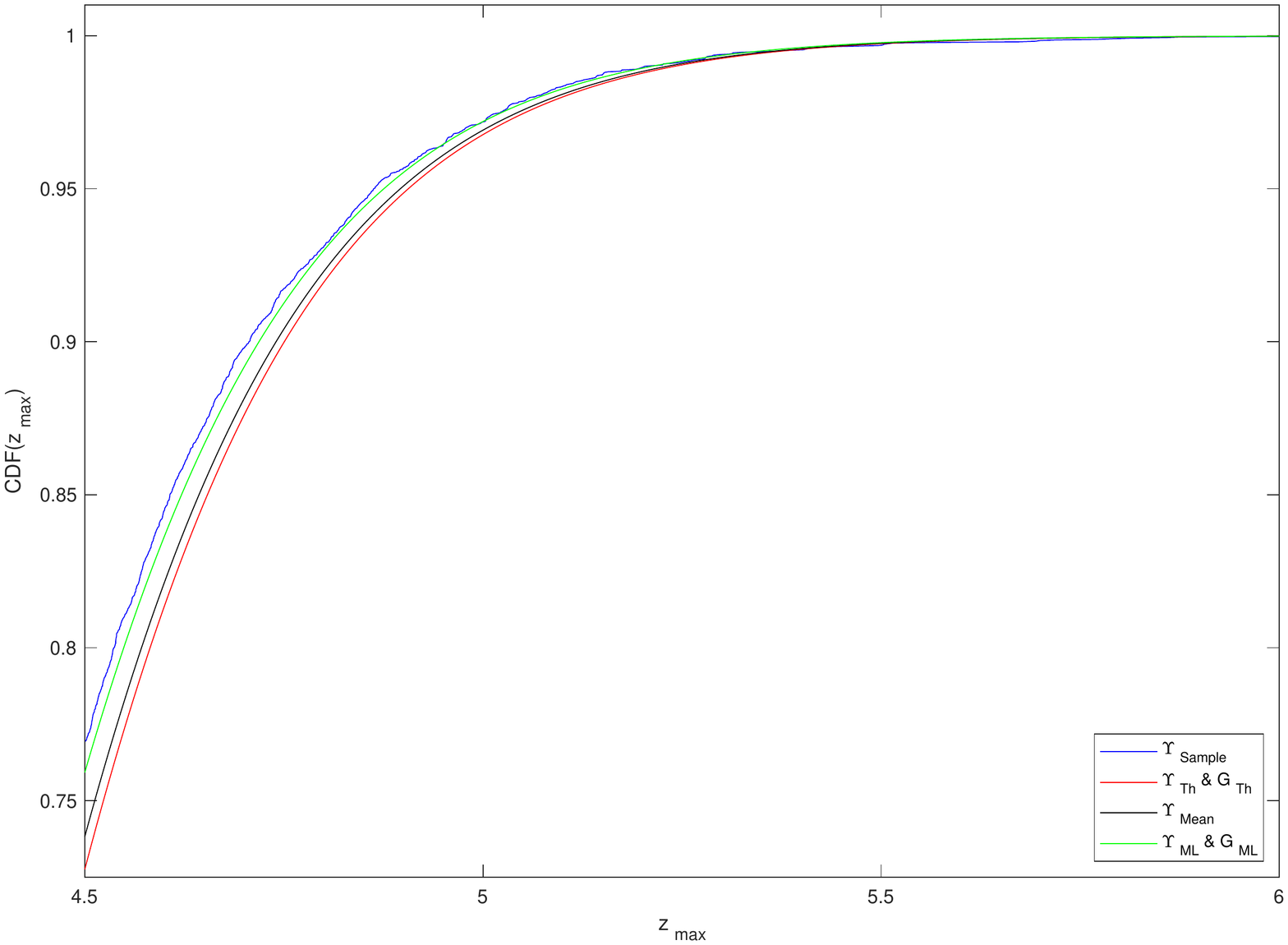}}
        \caption{Cumulative distribution functions corresponding to the PDFs in Fig.~\ref{fig:fig_pdf_circ}.}
        \label{fig:fig_cdf_circ}
    \end{figure}
\end{landscape}

Concerning the PAM, $\kappa=1$ and $N_p=2552$ are the theoretical values obtained as explained in VVA17, whereas 
the corresponding mean value of the maximum likelihood parameter $\hat{\kappa}$ and the number of peaks $N_p$ obtained from each of the simulated GRFs  are  $\bar{\kappa}=1.01 \pm 0.01$ and $\bar{N}_p = 2435 \pm 26$. 
The left panel of Fig.~\ref{fig:fig_pdf_circ} shows the PDFs $\upsilon_{\rm Th}(z_{\rm max})$ and $\upsilon_{\rm Mean}(z_{\rm max})$
corresponding to the theoretical value $N_p$ and the mean
value $\bar{N}_p$, respectively. Both slightly differ from the histogram $H(z_{\rm max})$. 
Given the small statistical fluctuation of $\bar{\kappa}$ and $\bar{N}_p$, such a discrepancy can be explained by the condition of complete spatial randomness assumed for the spatial distribution of the peaks.
In particular, the number of effectively iid peaks has to be expected to be smaller than both $N$ and $\bar{N}_p$. Indeed, a maximum likelihood fit $\upsilon_{\rm ML}(z_{\rm max})$  of the PDF given by Eq.~\eqref{eq:gz} to the set of simulated 
$\{ z_{\rm max} \}$ with $N_p$ as free parameter provides $\tilde{N}_p=2210$.
The right panel of Fig.~\ref{fig:fig_pdf_circ}  shows that something similar holds for the GDM since the theoretical value $N_*=17684$ is greater than the number $\tilde{N}_*=15330$ corresponding to the maximum likelihood fit
$g_{\rm ML}(z_{\rm max})$.
Here, it is worth stressing that $\upsilon_{\rm Th}(z_{\rm max})$ and $g_{\rm Th}(z_{\rm max}),$ and hence the corresponding CDFs $\Upsilon_{\rm Th}(z_{\rm max})$ and $G_{\rm Th}(z_{\rm max})$ shown in 
Fig.~\ref{fig:fig_cdf_circ} are almost identical. The same holds for 
$\upsilon_{\rm ML}(z_{\rm max})$ and $g_{\rm ML}(z_{\rm max})$ and the corresponding CDFs $\Upsilon_{\rm ML}(z_{\rm max})$ and $G_{\rm ML}(z_{\rm max})$ again shown in Fig.~\ref{fig:fig_cdf_circ}.
As expected the two approaches are practically equivalent, given that both methods are essentially based on the same assumptions. Moreover, in Fig.~\ref{fig:fig_cdf_circ} all these CDFs appear close to the sample CDF. This indicates
that the value of the parameters $N_p$ and $N_*$ is not of critical importance.

A further comparison of the two methods concerns the computational burden. In this respect the GDM is superior 
in situations where the autocorrelation function has circular symmetry (i.e., it is characterized by only a scale parameter), since the more expensive step consists in the computation of the quantities $\sigma_j$, $j=0,1,2$, which requires the numerical computation of three one-dimensional integrals. On the other hand, PAM requires the numerical computation of a number of integrals equal to the number of peaks selected for the detection. In general  the difference in computational 
time is on the order of a few seconds even in the case of large images. The computational superiority of GDM vanishes in the presence of autocorrelation functions when they are not circularly symmetric. This is because the three one-dimensional integrals become 
two-dimensional for a two-dimensional GRF and three-dimensional  for a three-dimensional GRF, which is computationally expensive. Finally, as shown in Eq.~\eqref{eq:sigma_i}, the GDM requires knowledge of the Fourier transform $w(k)$ of the filter used to 
obtain the GRF. If this is not available, it must be evaluated numerically with an additional computational cost.
In addition, the PAM is fully automatic given that the values of the parameters $\kappa$ and $N_p$ are estimated only from the peak amplitudes. Hence, contrary to the
GDM, there is no necessity to write a  specific code for a given autorrelation function.

\begin{landscape}
   \begin{figure}
        \resizebox{\hsize}{!}{\includegraphics{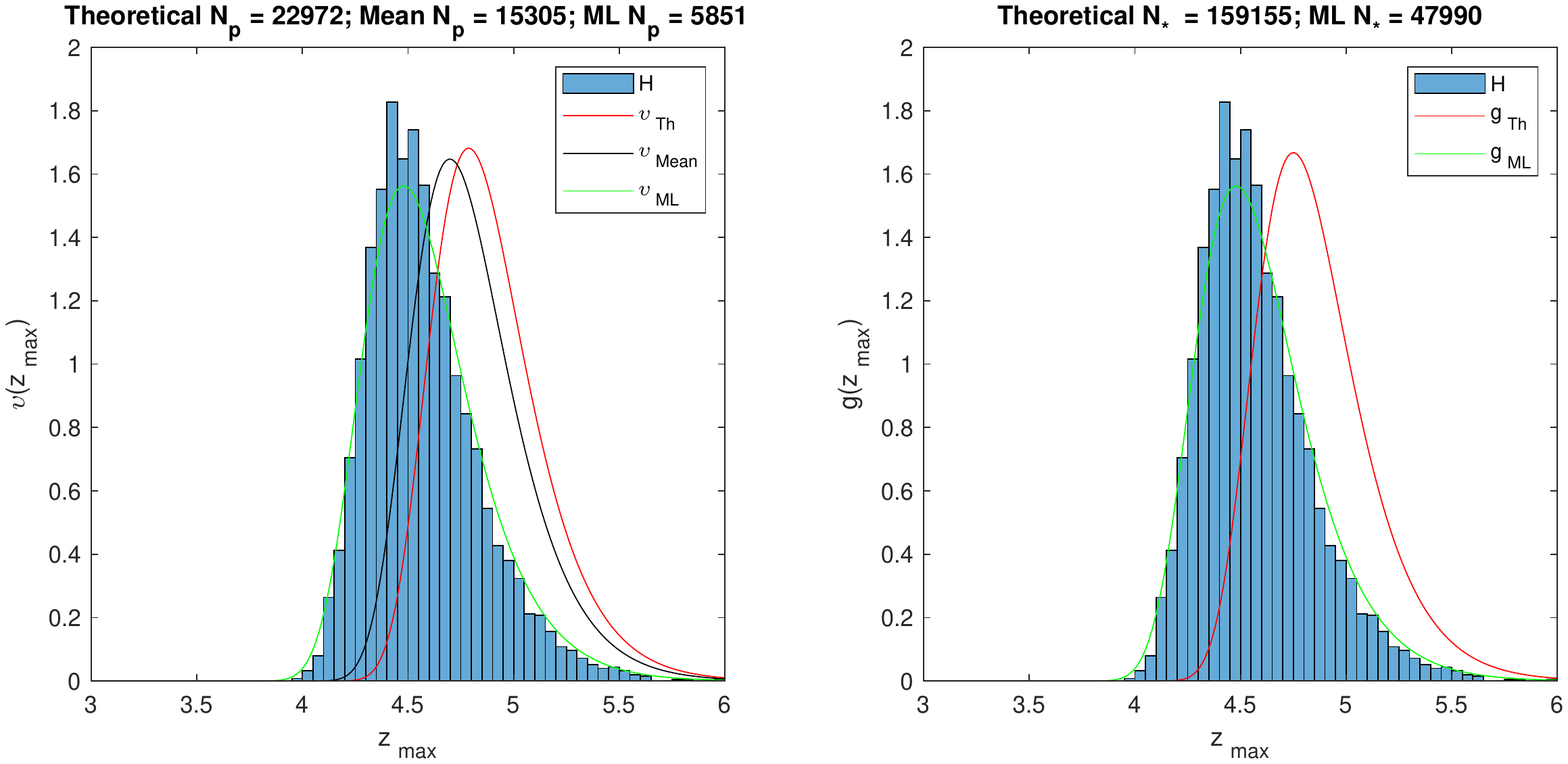}}
        \caption{As in Fig. \ref{fig:fig_pdf_circ} but with a circular Gaussian autocorrelation function with dispersion set to one pixel.}
        \label{fig:fig_pdf_circ_01_01}
    \end{figure}
\end{landscape}

\begin{landscape}
   \begin{figure}
        \resizebox{\hsize}{!}{\includegraphics{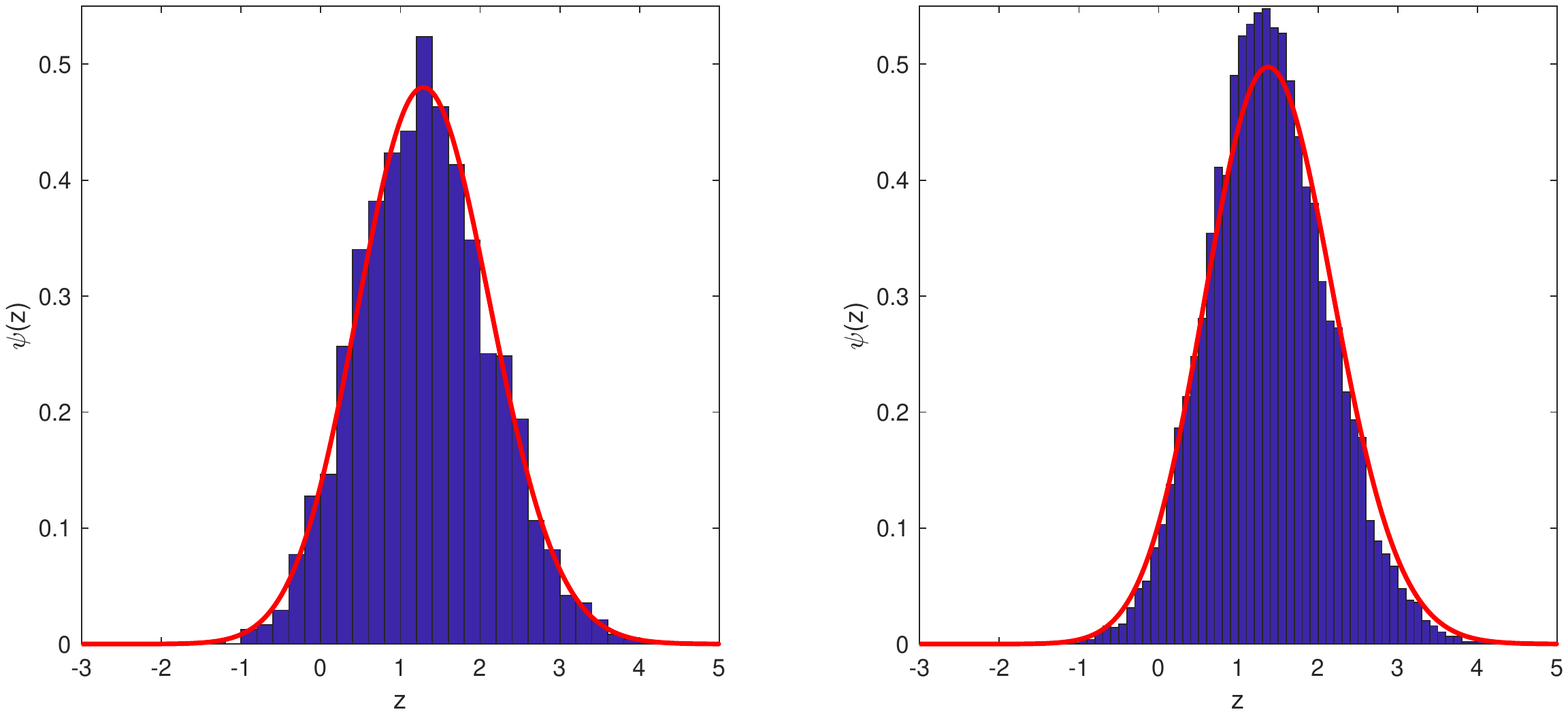}}
        \caption{Check for the applicability of the proposed detection procedure for two different situations (see text). Left panel: Histogram $H(z)$ vs. $\psi(z)$ from Eq.~\eqref{eq:pdf_z2}  for a numerical realization of a zero-mean unit-variance Gaussian random field with autocorrelation given by a circular Gaussian with dispersion set to three pixels. 
Right panel: As in the left panel but with the dispersion of the Gaussian autocorrelation function set to one pixel.}
        \label{fig:fig_check}
    \end{figure}
\end{landscape}

A final aspect to consider is the easiness with PAM to check if the PDF of the peak amplitudes corresponding to $\hat{\kappa}$ is compatible with the sample PDF.  This check is fundamental for the reliability of the detection. With GDM only the Gaussianity of the entries of the random field can be checked, hence the algorithm has to be used as a black box. In this respect, Fig.~\ref{fig:fig_pdf_circ_01_01} (to compare to Fig.~\ref{fig:fig_pdf_circ}) shows the result of a numerical experiment similar
to that presented above, but
where the dispersion of the circular Gaussian autocorrelation function is $\sigma_G=1$. It is clear that both PAM and GDM work really badly. The reason is that, again, both methods are developed in the context of a continuous GRF. With 
$\sigma_G=1$ the effects of the discretization become important. This point can be observed in Fig.~\ref{fig:fig_check}, which compares $\psi(z)$ for a typical realization of the GRF  when $\sigma_G=3$ (left panel) to that when 
$\sigma_G=1$ (right panel). As a one-sample Kolmogorov-Smirnvov test with a $99\%$ confidence level indicates, only when $\sigma_G=3$ is the sample distribution of the peak amplitudes compatible with the PDF expected for a GRF and hence only in this case can the two methods be safely used. The conclusion is that a blind application of both methods can lead to wrong results.

\section{Comparison with an ALMA map} \label{sec:ALMA}

\begin{landscape}
   \begin{figure}
        \resizebox{\hsize}{!}{\includegraphics{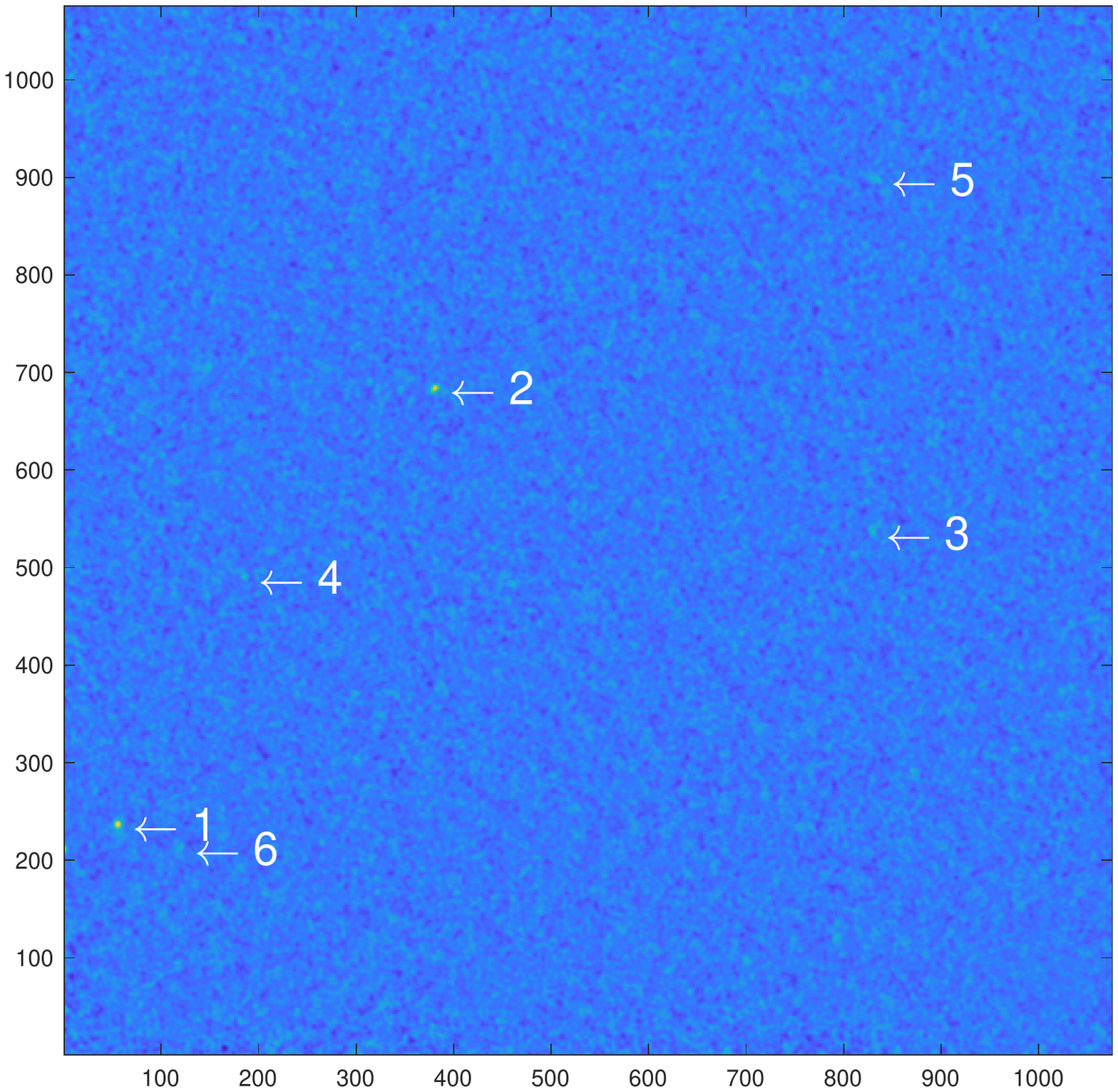}}
        \caption{Interferometric $1075 \times 1075$ pixels ALMA map used for testing the detection performances of PAM and GDM. Six point sources have been detected by both methods with a high level of confidence. 
They are labeled with a number in order of decreasing intensity. }
        \label{fig:fig_map}
    \end{figure}
\end{landscape}

The peak amplitudes method and GDM are here applied to the standardized zero-mean unit-variance interferometric ALMA map in Fig.~\ref{fig:fig_map} with the aim of detecting point sources.  The data are taken from the ALMA project 2012.1.00173.S, a mosaicing of the Hubble Ultra-Deep Field (HUDF) in continuum \citep{dun17}. The HUDF was observed using a 45-pointing mosaic, with each pointing separated by 0.8 times the antenna beamsize. The details of the observing and data reduction procedures can be found in Dunlop et al. (2017). For the purpose of this work, in order to simplify the analysis, we have cut the outer edges of the total image and produced a symmetric, square-shaped image of size $1075 \times 1075$ pixels
(corresponding to $110\times 110$ arcsec$^2$). On this inner part we applied our algorithm and found $11959$ peaks.

As explained in Vio16, in ALMA maps the point sources and the blob-shaped structures due to the noise have a similar aspect and the MF cannot be applied. Consequently,
the detection test becomes a thresholding test, where a peak in the map is assigned to a point source if it exceeds a given threshold.
Before applying the detection algorithm, it is necessary to check whether the map meets the requirements mentioned above. In particular, the Gaussianity and isotropy of the noise background, the compatibility of the sample  PDF of the peak amplitudes
with the theoretical PDF $\psi(z),$  and the fact that the spatial distribution of the peaks is compatible with a CSRPP.

\begin{landscape}
   \begin{figure}
        \resizebox{\hsize}{!}{\includegraphics{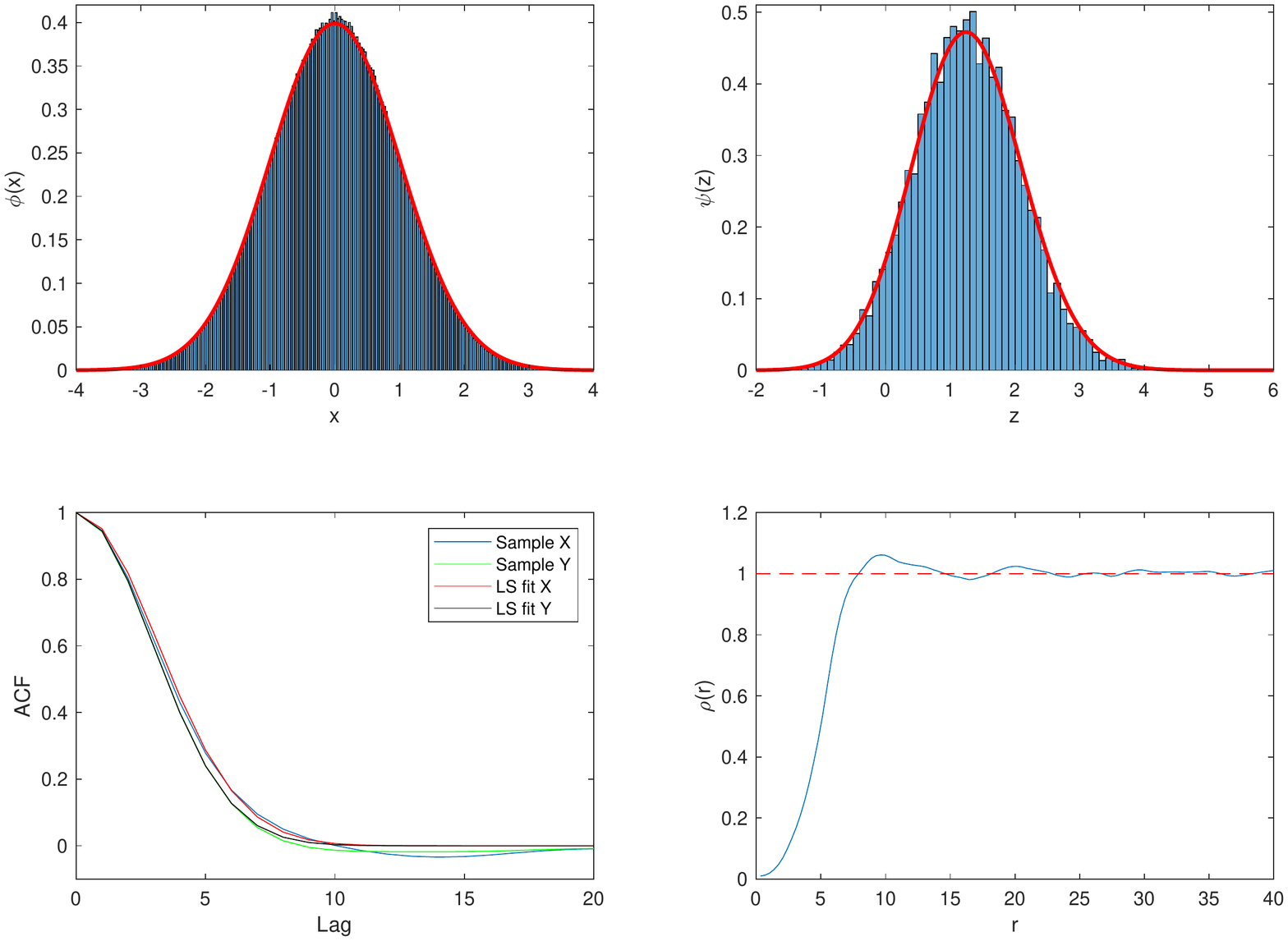}}
        \caption{Checks of the conditions of applicability of the detection procedure for the interferometric ALMA map (see text). Top-left panel: Histogram of the values of the pixels vs. the standard Gaussian PDF $\phi(x)$. 
Top-right panel: Histogram of the peak values vs. the theoretical PDF 
$\psi(z)$ given by Eq.~\eqref{eq:pdf_z2}. Bottom-left panel: Slices along the $X$ and the $Y$ directions of the sample autocorrelation function vs. the corresponding slices of a least-square fit with a two-dimensional Gaussian function.
Bottom-right panel: Sample pair correlation function $\rho(r)$ of the peaks. The red line provides the theoretical $\rho(r)$ due to a CSRPP.}
        \label{fig:fig_test}
    \end{figure}
\end{landscape}

The results of these checks are presented in  Fig.~\ref{fig:fig_test}. In particular, the top-left panel shows that, although similar to a Gaussian, the histogram of the entries of the map indicates a PDF slightly leptokurtik\footnote{A leptokurtik PDF has a shape more peaked than a Gaussian.}.
The excess of values close to the mean can be understood taking into account that in general a zero-mean, isotropic GFR with a smooth autocorrelation function shows a lack of entries with high absolute value\footnote{This can be understood taking into account that an isotropic GFR with a specific autocorrelation function can be obtained by filtering a Gaussian white-noise process by means of a linear filter.
In most of the astronomical applications, the filters necessary to obtain a specific autocorrelation function are of low-pass type (e.g., for a Gaussian autocorrelation function the filter is also a Gaussian). As a consequence, in the resulting GRF the greatest values of the white-noise process are smoothed out in favor of values closer to the mean. As a matter of fact, this effect becomes important only for the maps 
with dimensions comparable to the extension of the area where the autocorrelation function is significantly different from zero, but this is not the case for the map under consideration.}.
More importantly, from the top-right panel of the same
figure the histogram of the peak amplitudes appears compatible with the $\psi(z)$ corresponding  to a maximum likelihood estimate $\hat{\kappa}=0.98$. This is confirmed by a Kolmogorov-Smirnov test at a confidence level of $99\%$,

The isotropy of the noise background is supported by the bottom-left panel of Fig.~\ref{fig:fig_test}, which displays the sample autocorrelation functions along the $X$ and $Y$
directions versus the corresponding slices of a two-dimensional circular Gaussian resulting from a least-squares fit. The good agreement indicates that the ALMA map is compatible with an isotropic Gaussian GRF 
whose autocorrelation function is a two-dimensional circular Gaussian with dispersion $\sigma_G \approx 3$.

Finally, as is visible in the bottom-right panel of the same figure, the sample pair correlation function $\rho(r)$ is compatible with a CSRPP for $r \ge 7$.
Since $\varrho(7) \approx 0,$ this means that, as seen in Sect.~\ref{sec:comparison}, the peak amplitudes do not present a relevant mutual statistical dependence, hence most of them can be considered iid.

\begin{landscape}
   \begin{figure}
        \resizebox{\hsize}{!}{\includegraphics{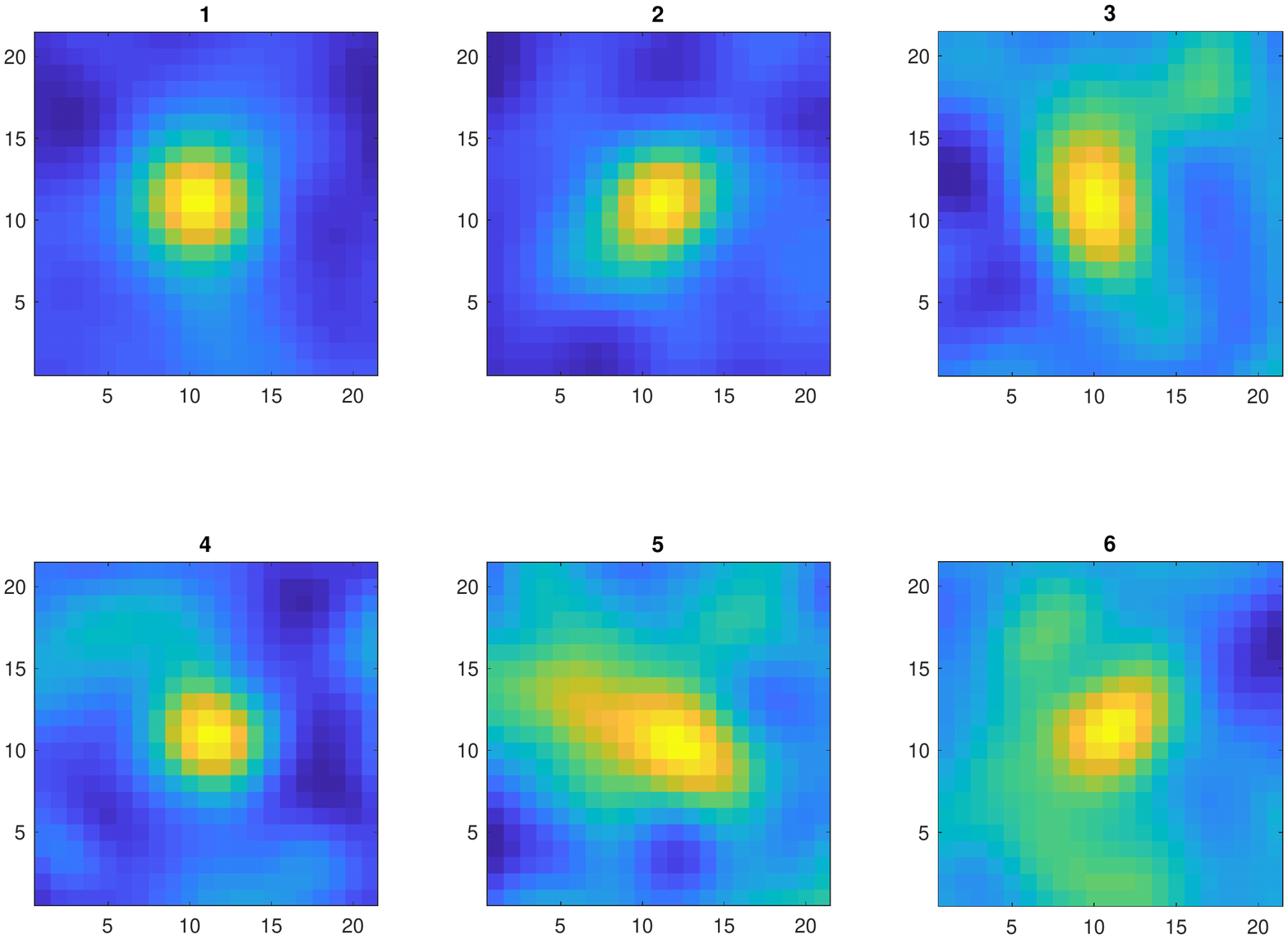}}
        \caption{Sub-maps corresponding to the areas of the point sources detected in the map in Fig.~\ref{fig:fig_map}.}
        \label{fig:fig_submaps}
    \end{figure}
\end{landscape}

Both PAM and GDM produce six statistically significant  detections, which are labeled in Fig.~\ref{fig:fig_map} with a number in order of decreasing intensity and shown in more detail in Fig.~\ref{fig:fig_submaps}. The SPFAs coming from the two methods are very similar, since for PAM the values are $0$, $4.65\Erm-11$, $1.51\Erm-06$, $5.76\Erm-06$, $3.68\Erm-02$, $7.05\Erm-02$, whereas for the GDM they become $0$, $0$, $1.54\Erm-06$, $5.84\Erm-06$, $3.74\Erm-02$, $7.16\Erm-02$. 
This result confirms the equivalence of the two methods as it concerns the detection performance. Table~\ref{tab:sources} reports the coordinates of the detected sources and their identification with the sources reported in \citet{dun17}.

\begin{table*}
\caption{Identification of the sources on the map in Fig.~\ref{fig:fig_map}.}          
\label{tab:sources}      
\hskip -0.7cm          
\begin{tabular}{c c c c}     
\hline\hline       
\\
Source number  & identification & coordinates & \\ 
& & RA & DEC\\

\hline    
\\

Source 1 &   UDF2 & 3h32m43.53s &  -27d46$^\prime$39.25$^{\prime\prime}$ \\
Source 2 &   UDF3 & 3h32m38.55s &  -27d46$^\prime$34.57$^{\prime\prime}$ \\
Source 3 &   UDF5 & 3h32m36.96s &  -27d47$^\prime$27.13$^{\prime\prime}$ \\
Source 4 &   UDF4 & 3h32m41.02s &  -27d46$^\prime$31.58$^{\prime\prime}$ \\
Source 5 &   UDF6 & 3h32m34.43s &  -27d46$^\prime$59.72$^{\prime\prime}$ \\
Source 6 &  NOT IDENTIFIED &3h32m38.66s &  -27d48$^\prime$06.12$^{\prime\prime}$ \\ 

\\
\hline           
    
\end{tabular}
\end{table*}

\section{Conclusions}  \label{sec:conclusions}

In this paper the performances of two techniques, the  PDF of the peak amplitudes method (PAM) and the Gumbel distribution method (GDM), have been compared in the context of the detection of weak signal embedded in noise. 
The two methods have been applied to simulated signals and to observations taken with the ALMA interferometer.
We have shown that the two approaches are almost perfectly equivalent in their detection capability, but PAM proves to be more flexible and, at the same time, allows for an easy control of the condition of applicability of the technique. Hence, it appears
more appropriate in the search for weak sources in observations dominated by noise.

\begin{acknowledgements}

\end{acknowledgements}

\clearpage

\clearpage
\clearpage
\clearpage
\clearpage
\clearpage
\clearpage
\clearpage

\end{document}